\documentclass[12pt,preprint]{aastex}
\usepackage{emulateapj5}
\usepackage{epsfig}

\newcommand{\etal}{\mbox{et al.}}
\newcommand{\ergcms}{erg cm$^{-2}$ s$^{-1}$}
\newcommand{\msunyr}{M$_\odot$ y$^{-1}$}
\newcommand{\ergsec}{erg s$^{-1}$}

\newcommand{\degree}{$^\circ$}

\newcommand{\chandra}{{\it Chandra}}

\newcommand{\granat}{{\it GRANAT}}
\newcommand{\asca}{{\it ASCA}}
\newcommand{\xmm}{{\it XMM-Newton}}
\newcommand{\rxte}{{\it RXTE}}

\newcommand{\bepposax}{{\it BeppoSAX}}

\newcommand{\sgrastar}{\mbox{Sgr A$^*$}}

\newcommand{\program}[1]{{\tt {#1}}}
\newcommand{\html}[1]{{\tt http://#1}}

\newcommand{\gcllb}{\mbox{GRS~1741.9$-$2853}}

\makeatletter
\newenvironment{inlinefigure}{%
\def\@captype{figure}%
\noindent\begin{minipage}{0.999\linewidth}\begin{center}}
{\end{center}\end{minipage}\smallskip}
\makeatother

\shortauthors{Muno \etal}
\shorttitle{\gcllb}

\submitted{Submitted to ApJ Letters}
\begin{document}
\title{The X-ray Binary \gcllb\ in Outburst and Quiescence}
\author{M. P. Muno, F. K. Baganoff, and J. S. Arabadjis}

\affil{Center for Space Research,
Massachusetts Institute of Technology, Cambridge, MA 02139;
muno@space.mit.edu}

\begin{abstract}
We report \chandra\ and \xmm\ observations of the transient neutron star 
low-mass
X-ray binary \gcllb. \chandra\ detected the source in outburst on 2000 
October 26 at an X-ray luminosity of $\sim 10^{36}$~\ergsec\ (2--8~keV; 8 kpc),
and in quiescence on 2001 July 18 at $\sim 10^{32}$~\ergsec. 
The latter observation is the first detection of \gcllb\ in quiescence.
We obtain an accurate position for the source of 
17h 45m 2.33s, -28\degree 54\arcmin 49.7\arcsec\ (J2000), with an 
uncertainty of 0.7\arcsec. \gcllb\ was not detected significantly in
three other \chandra\ observations, nor in three \xmm\ observations, 
indicating that the luminosity of the source in quiescence varies by at 
least a factor of 5 between $(< 0.9 - 5.0) \times 10^{32}$~\ergsec\ 
(2--8 keV). 
A weak X-ray burst with a peak luminosity of $5 \times 10^{36}$~\ergsec\
above the persistent level was observed with \chandra\ during the outburst 
on 2000 October 26. The
energy of this burst, $10^{38}$~erg, is unexpectedly low, and may suggest
that the accreted material is confined to the polar caps of the neutron
star. A search of the literature reveals that \gcllb\ was observed in 
outburst with \asca\ in Fall 1996 as well, when the \bepposax\ WFC 
detected the three previous X-ray bursts from this source. The lack of X-ray
bursts from \gcllb\ at other epochs suggests that it produces bursts
only during transient outbursts when the accretion rate onto the surface of 
the neutron star is about $10^{-10}$~\msunyr. A similar situation may hold 
for other low-luminosity bursters recently identified from WFC data. 
\end{abstract}

\keywords{stars: individual: \gcllb --- stars: neutron --- X-rays: bursts}

\section{Introduction\label{sec:intro}}
In recent years, the \bepposax\ Wide-Field Camera (WFC) and the 
{\it Rossi X-ray Timing 
Explorer} All-Sky Monitor (\rxte\ ASM) have identified two
classes of faint low-mass X-ray binary (LMXB) that may be closely
related: low-luminosity transients and low-luminosity bursters. 
The low-luminosity transients consist of a rather inhomogeneous group of 
about 15 LMXBs with 
outbursts that last from a few days to several months, and get no brighter 
than a few times $10^{36}$~\ergsec, which corresponds to an accretion rate 
of $10^{-10}$ \msunyr\ \citep[e.g.,][]{zand00}. 
This distinguishes the low-luminosity transients
from more easily detected transient LMXBs such as Aql~X-1 
%and XTE~J1550$-$564 \citep{bel02} 
that usually exhibit outbursts brighter
than $10^{37}$~\ergsec, and from which faint outbursts are less
common \citep[e.g.,][]{sim02}. The faintness of these outbursts has been 
attributed
to average mass transfer rates of $\dot{M} \lesssim 10^{-11}$ \msunyr\
\citep{zand00, king00}. The low-luminosity transients have attracted 
particular attention because they include the four known 
accreting millisecond X-ray pulsars \citep{wk98,mar02,gal02,mar03}. For this
reason, it has been hypothesized that 
the low average accretion rates allow relatively strong ($> 10^{8}$~Gauss) 
magnetic fields to persist on the surfaces of the neutron stars among these
LMXBs, whereas the surface field is buried in systems with higher accretion 
rates (Cumming, Zweibel, \& Bildsten 2001)\nocite{cum01}.

The low-luminosity X-ray bursters are sources from which bright thermonuclear
X-ray bursts 
(see Lewin, van Paradijs, \& Taam 1993 for a review)\nocite{lvt93} have
been observed with the \bepposax\ WFC, and yet there was no evidence for 
X-ray emission from persistent accretion at the time of the burst 
\citep{coc01,cor02a,cor02b}. 
This sample of low-luminosity bursters may represent a large population of 
undiscovered neutron star X-ray binaries, depending upon how often these
systems produce bursts. The intervals between X-ray bursts is a strong function
of the accretion rate per unit area onto the neutron star, which determines 
how quickly a sufficient column of material is collected for helium burning 
to become unstable \citep[e.g.,][]{bil00}. Since the nuclear energy 
in accreted material is at most 4\% of the gravitational energy that 
it emitted during accretion \citep{lvt93}, sufficient nuclear 
energy to produce an easily-detectable $10^{39}$~erg burst is collected 
in 5 hours on the surface of a neutron star producing $10^{35}$~\ergsec\ of 
X-rays through accretion (corresponding to an accretion rate of $10^{-11}$ 
\msunyr), 
while such a burst could not occur for several years on a star emitting 
persistently at $10^{32}$~\ergsec\ ($10^{-14}$ \msunyr). Unfortunately, 
the {\it BeppoSAX} WFC could only 
place an upper limit of $10^{36}$~\ergsec\ on the luminosity of persistently
faint bursters close to the Galactic center, so estimates of the frequency of 
bursts from 
the low-luminosity bursters are uncertain by several orders of magnitude. 
Therefore, the number of low-luminosity bursters can only be guessed within 
a factor of 1000, and it is still possible that many of them are also
faint X-ray transients.

\gcllb\ is both a faint transient system and a low-luminosity burster. It
was discovered as a transient X-ray source located 10\arcmin\ from the 
Galactic Center with {\it Granat}, during two observations 
on 1990 March 24 and April 8 (Pavlinsky, Grebenev, \& Sunyaev 1994). 
The flux during these observations remained 
approximately
%%%
\begin{figure*}[ht]
\epsscale{0.8}
\centerline{\epsfig{file=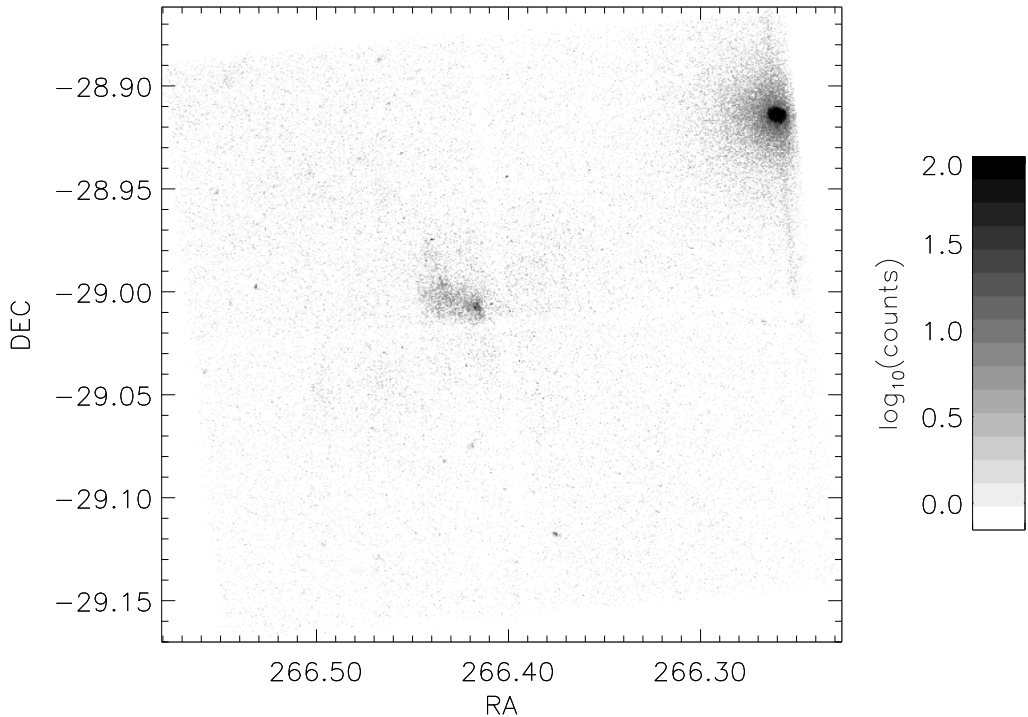,width=0.75\linewidth}}
\caption{Image in the 0.5-8.0~keV band of the ACIS-I field containing 
\gcllb\ in outburst on 2000 October 26. \gcllb\ lies in the upper right.
The bright region at the center is the Sgr A complex.}
\label{fig:obsid1561a}
\end{figure*}
%%%
 constant at $2 \times 10^{-10}$~\ergcms\ (4--30~keV), but 
by the Fall of 1990 the source had decreased in flux 
by at least a factor of seven, making it undetectable with {\it Granat}.
In the Fall of 1996, the \bepposax\ Wide-Field 
Camera (WFC) detected three thermonuclear X-ray bursts from the source, 
identifying the compact object in \gcllb\ as a weakly-magnetized
neutron star \citep{coc99}. The brightest of 
these bursts had a peak flux of $4 \times 10^{-8}$~\ergcms, and exhibited 
a ``radius expansion'' episode in which the photosphere of the neutron star 
appeared to be driven from its surface by radiation 
pressure. Assuming that the peak intensity of this burst represents the 
Eddington limit for a neutron star with standard parameters 
($\approx 2 \times 10^{38}$ \ergsec; see Lewin \etal\ 1993), the
system lies near the Galactic Center, at a distance of approximately
8~kpc (since \gcllb\ also is only 10\arcmin\ away from the Galactic 
center, we adopt this as its distance throughout the paper). At this 
distance, the luminosities of the transient outbursts were only 
$\approx 10^{36}$ \ergsec. Repeated observations of the Galactic center
by the \bepposax\ WFC each Spring and Fall from mid-1996 through 2001 did not 
detect any outbursts from \gcllb\ above a sensitivity limit corresponding to 
$3 \times 10^{36}$~\ergsec\ (2--28~keV) \citep{cor02a}, and 
continuous monitoring with the \rxte\ All-Sky Monitor rules out 
outbursts brighter than $6 \times 10^{36}$~\ergsec\ (2--12~keV) lasting more
than a month from 1996 until the present. However, the source was detected 
twice as part of the \asca\ survey of the Galactic Plane, with luminosities 
between $10^{34}$ and $10^{36}$~\ergsec\ (1--10~keV) 
\citep[see Table~\ref{tab:fluxhistory} and][]{sak02}.

In this paper, we report several serendipitous \chandra\ and 
\xmm\ observations of \gcllb. \chandra\ detected the source in outburst, 
which allowed us to determine its location precisely \citep{mun03}. 
Using this location, we study the flux from this source in quiescence. We 
also report the detection of a thermonuclear X-ray burst while
the source was in outburst. 

\section{Observations and Data Analysis}

\subsection{\chandra}

To date, the location of \gcllb\ has been observed on five occasions with 
\chandra:
three times as part of a program to monitor the super-massive black hole
at the center of the Galaxy, \sgrastar\ \citep{bag01, bag03, mun03}, and
twice as part of a shallow survey of the inner 300~pc of the Galaxy
\citep{wang02}. The observations are listed in Table~\ref{tab:obs}.

All five observations were taken using the 
Advanced CCD Imaging Spectrometer imaging array (ACIS-I). 
%\citep[ACIS-I][]{gar02}.
The ACIS-I is a set of four 1024-by-1024 pixel CCDs, covering
a field of view of 17\arcmin\ by 17\arcmin. When placed on-axis at the focal
plane of the grazing-incidence X-ray mirrors, the imaging resolution 
is determined by the pixel size of the CCDs, 0\farcs492. The CCDs 
measure the energy of incident photons with a resolution of 50-300 eV 
(depending on photon energy and distance from the read-out node) within a 
calibrated energy band of 0.5--8~keV.

We reduced the data starting with the level 1 event files provided by 
the \chandra\ X-ray Center (CXC). We first removed the pixel randomization 
applied by the default processing software, and corrected the pulse 
heights of each event for the position-dependent charge-transfer inefficiency 
caused by radiation damage early in the mission \citep{tow00}. 
We excluded most events flagged as possible non-X-ray background, but 
left in possible 
cosmic ray after-glows because they are difficult to
distinguish from 
the strong diffuse emission and numerous point 
sources near the Galactic center. We applied the standard \asca\ grade filters 
to the events, as well as the good-time filters supplied by the CXC. We 
examined each observation for background flares, and removed 10 ks of 
strong flaring from ObsID 0242 when the count rate rose to 3 standard 
deviations above the mean. Finally, we corrected the 
pointing aspect for each observation as specified in the \chandra\ data
caveats 
page.\footnote{http://cxc.harvard.edu/cal/ASPECT/fix\_offset/fix\_offset.cgi}
%Exposure maps were created 
%assuming a monochromatic incident spectrum with a photon energy of 3~keV,
%in order to keep track of regions from which other point sources were removed.

To search for point sources in each observation, we employed the 
CIAO tool \program{wavdetect} \citep{fre02} in the manner described in 
\citet{mun03}. We set the detection threshold to $10^{-7}$,
which corresponds to the probability that a background fluctuation in any given
pixel will be identified as a source. 
The resulting source list for the \sgrastar\ monitoring campaign, which covers
ObsIDs 0242, 1561a, and 1561b, is presented 
in \citet{mun03}. The source lists for ObsIDs 2284 and 2287 will be reported 
elsewhere. These lists were used to identify candidate counterparts to 
\gcllb\, and to exclude point sources from the estimate of the background flux.
We verified that the astrometry 
for each observation was accurate to within 0\farcs5 by cross-correlating
the positions of the half-dozen sources that were both detected in 
the soft X-ray band (0.5--1.5~keV) and present in the USNO-B 
catalog.\footnote{D. Monet \etal; see 
\html{www.nofs.navy.mil/projects/pmm/catalogs.html}}

\gcllb\ was evident by eye as a bright transient in the image from ObsID 1561a 
(2000 October 26; Figure~\ref{fig:obsid1561a}). Since the source is located 
10\arcmin\ from the aim-point of the observation, its location is uncertain 
by about 5\arcsec\ \citep[see also][]{mun03}. Its position is within the 
1\arcmin\ error circles of previous localizations of \gcllb\ 
with \granat\ and \asca. The transient source exhibited a 
thermonuclear burst during this \chandra\ observation (see below), so 
we can confidently associate it with \gcllb. We searched the 
source lists from the other observations, and identified a faint point 
source within our initial localization of \gcllb\ at 6\arcmin\
from the aim point of ObsID 2287. Its position was 
17h 45m 2.33s, -28\degree 54\arcmin 49.7\arcsec\ (J2000), with an uncertainty
of 0.7\arcsec\ \citep[compare][]{bra01}.
No point source was identified by \program{wavdetect} 
within 5\arcsec\ of \gcllb\ in the other three observations.

Having established a precise position for \gcllb, we estimated its flux
in each of the five 
observations using the \program{acis\_extract} routine from the Tools for 
X-ray Analysis \citep{bro02}. We extracted event lists for the source 
from a polygonal region chosen 
to match the contour of 90\% encircled energy from the point-spread function 
(PSF) at an energy 
of 4.5~keV. Background event lists were extracted from circular regions 
centered on \gcllb, excluding from the event lists counts in circles 
circumscribing the 95\% contours of the PSF around the point sources detected 
in each field. 
The number of total, background, and net counts from each observation
are listed in Table~\ref{tab:counts}. 
We computed 

\begin{inlinefigure}
\epsscale{0.8}
\centerline{\epsfig{file=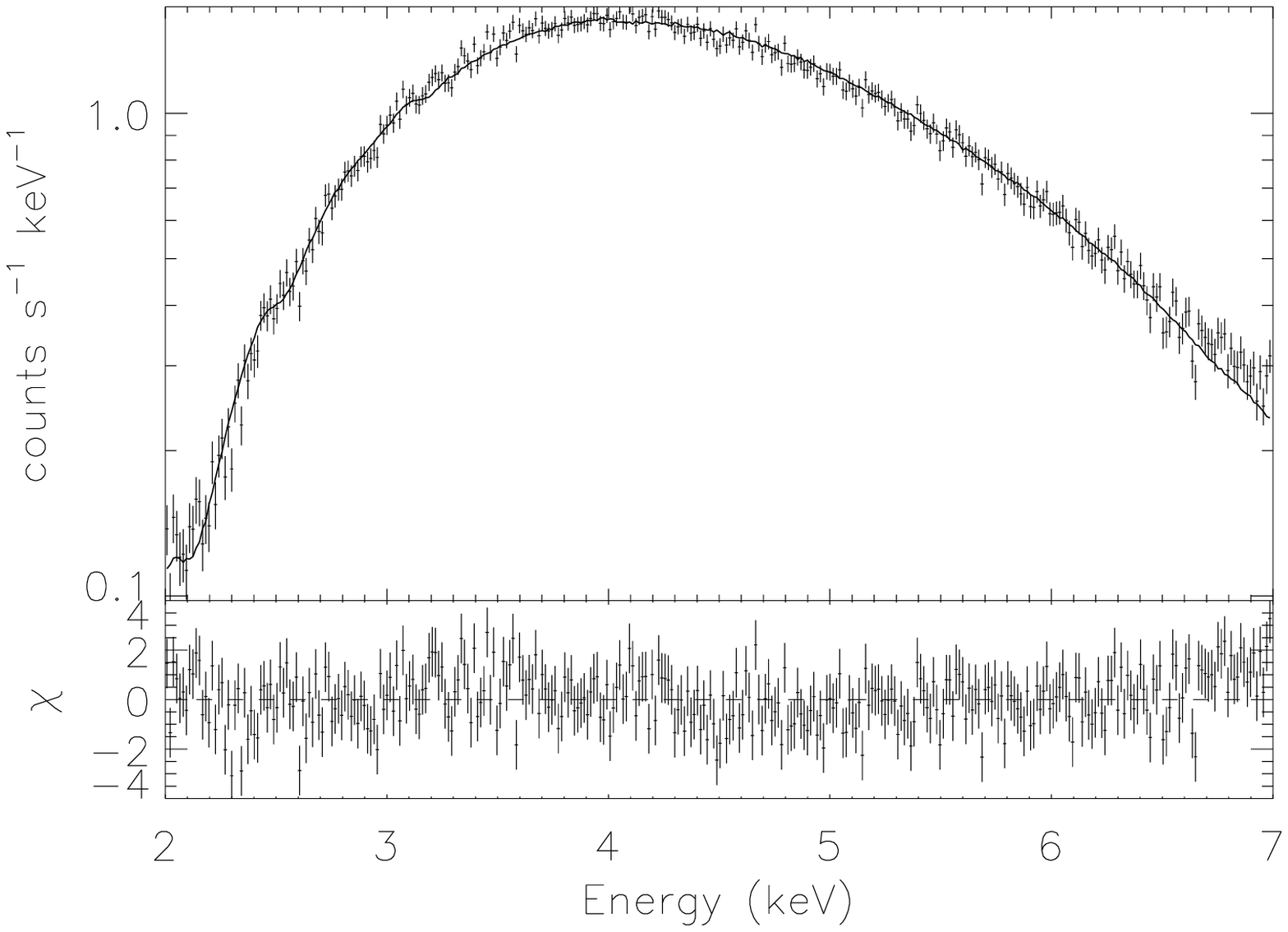,width=\linewidth}}
\caption{ACIS-I spectrum of \gcllb\ while in outburst on 2000 October 26.
The {\it top panel} displays the number of counts received as a function
of energy, which depends both on the incident spectrum and the detector
effective area. 
The spectrum is modeled with an absorbed power law continuum (solid line). 
The {\it bottom panel} displays the residuals after applying the model, 
normalized to the 1$\sigma$ uncertainty on the counts in each spectral bin.}
\label{fig:spec}
\end{inlinefigure}

\noindent
90\% uncertainties and upper limits on the 
net number of counts using the expected Poisson
distribution of the total number of counts given the background derived from
the photometry \citep{kbn91}. 

We extracted a spectrum for ObsID 1561a (2000 October 26), which contained 
170,214 counts from 35,705 s of live time in an extraction region of 
1563 pixels. 
We used response matrices appropriate for data with the charge-transfer 
inefficiency corrected, and an effective area function (ARF)
corrected to account for the fraction of the PSF enclosed by the extraction 
region and for the hydrocarbon build-up on the 
detectors.\footnote{\html{cxc.harvard.edu/cal/Acis/Cal\_prods/qeDeg/}}
The burst contributed only 250 counts to the spectrum, so we did not 
remove it from the persistent spectrum.
Although a count rate this high (5 count s$^{-1}$) would generally cause 
serious pile-up if the source were observed on-axis with {\it Chandra}, 
the 10\arcmin\ offset angle spread the counts over a large enough 
area on the ACIS-I CCDs that pile-up is not an issue. The mean count rate
per pixel is only 0.003 count s$^{-1}$ pixel$^{-1}$, which produces a 
pile-up fraction of 0.3\%. The brightest pixel only contained 803 
counts, for a rate of 0.02 count s$^{-1}$ pixel$^{-1}$ and a pile-up fraction
of 2\%. 

We fit the spectrum between 2--7~keV, because outside this energy range there 
were systematic residuals. These residuals may have been caused by the 
high count rate incident on a relatively small area of the CCD, which may 
have changed the efficiency of charge-transfer as the events were read out.
We modeled the spectrum with a power-law continuum absorbed and scattered by 
interstellar gas and dust.\footnote{We modified the dust model from the 
default version in \program{XSPEC} so that it did not assume that the dust 
was optically thin.} We linked the optical depth of dust to 
the absorption using the relation 
$\tau = 0.485 \cdot N_{\rm H}/(10^{22} {\rm cm}^{-2})$ \citep{bag01},
and fixed the halo size to 100 times that of the PSF. The fit had a 
$\chi^2/\nu = 403/339$, and is displayed in Figure~\ref{fig:spec}.
The best-fit photon index was $\Gamma = 1.88 \pm 0.04$ and the column density 
$N_{\rm H} = (9.7 \pm 0.2) \times 10^{22}$

\begin{inlinefigure}
\epsscale{0.8}
\centerline{\epsfig{file=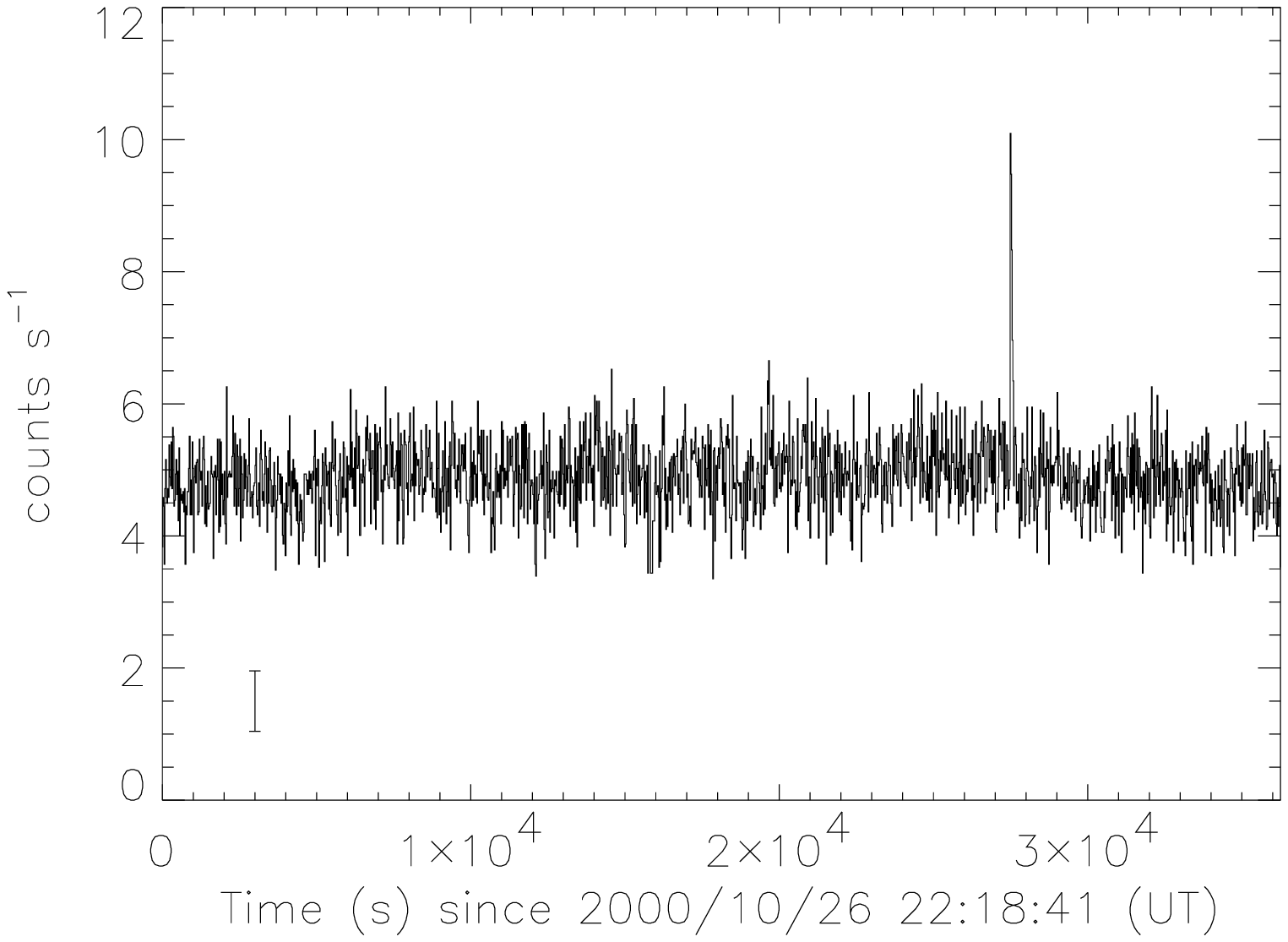,width=\linewidth}}
\caption{Light-curve of \gcllb\ in outburst in the 0.5-8.0~keV energy
band. An X-ray burst lasting about 100~s is observed 25,700 s into the 
observation. A typical
error bar is displayed in the lower left.}
\label{fig:lc}
\end{inlinefigure}

\noindent
cm$^{-2}$. 
We extrapolated the fit to compute an ``observed'' X-ray flux of 
$F_{\rm X} = 2 \times 10^{-10}$~\ergcms between 2--8 keV, where we 
chose the energy range to be consistent with the
catalog in \citet{mun03}.\footnote{The lower bound of 2~keV was chosen because 
Galactic absorption limits the flux observed below this value. The upper bound
of 8~keV is the limit above which the ACIS effective area is small, and its
spectral response is not well calibrated.} The deabsorbed X-ray
flux between 2--8~keV was $4 \times 10^{-10}$~\ergcms, which corresponds to a
luminosity
of $3 \times 10^{36}$ erg s$^{-1}$ at 8 kpc. 

For the remaining four observations, we converted the count rates and 
upper limits into fluxes using \program{PIMMS}, and list the results in 
Table~\ref{tab:counts}. The source was detected significantly ($> 7\sigma$)
in ObsID 2287 on 2001 July 18, and marginally in ObsID 1561b on 2001 July 14
($2\sigma$) and in ObsID 2284 on 2001 July 18 ($4\sigma$). If 
we assume that the spectrum is the same as in outburst, the observed flux
was $(1-5) \times 10^{-14}$ \ergcms\ 
(2--8~keV). Assuming a 0.3~keV blackbody typical for 
a quiescent neutron star LMXB (Brown, Bildsten, \& Rutledge 1998) 
results in fluxes a factor of 3 lower. 
Including the uncertainty on the spectrum in quiescence,
the deabsorbed luminosities range between 
$(1-10) \times 10^{32}$~erg s$^{-1}$ (2--8~keV, 8 kpc), which is typical 
for a LMXB in quiescence. There is evidence for variability in the
quiescent flux, as the upper limit on 1999 September 21 is lower than the 
detection on 2001 July 18 at 17:38 with a significance greater than
3-$\sigma$. 
%with a significance of $7 \times 10^{-4}$.

We extracted a light curve from the observation in which the 
transient was bright (ObsID 1561a on 2000 October 26), 
which we display in Figure~\ref{fig:lc}. The bin size is 23~s, so
that there are over 100 counts per bin. An X-ray burst lasting about 100~s 
is observed 25,500~s into the observation (Figure~\ref{fig:burst}a). It 
decays exponentially on a time scale of $54 \pm 6$~s. We examined the
burst in two energy bands, 0.5--4.5~keV and 4.5--8.0~keV, which provide 
approximately equal counts during 
the persistent periods before and after the burst. The burst is harder 
in the peak than in the
tail (Figure~\ref{fig:burst}b), and the profile decays faster at high 
energies ($\tau = 43 \pm 5$~s) than at low energies ($\tau = 77 \pm 18$~s). 
Thus, the burst appears to cool as it decays, which suggests that the burst 
is thermonuclear in origin, and not a sudden burst of accretion
\citep[see][]{lvt93}. However, the burst 
is weak, reaching a peak flux only 3.5 times that of

\begin{inlinefigure}
\epsscale{0.8}
\centerline{\epsfig{file=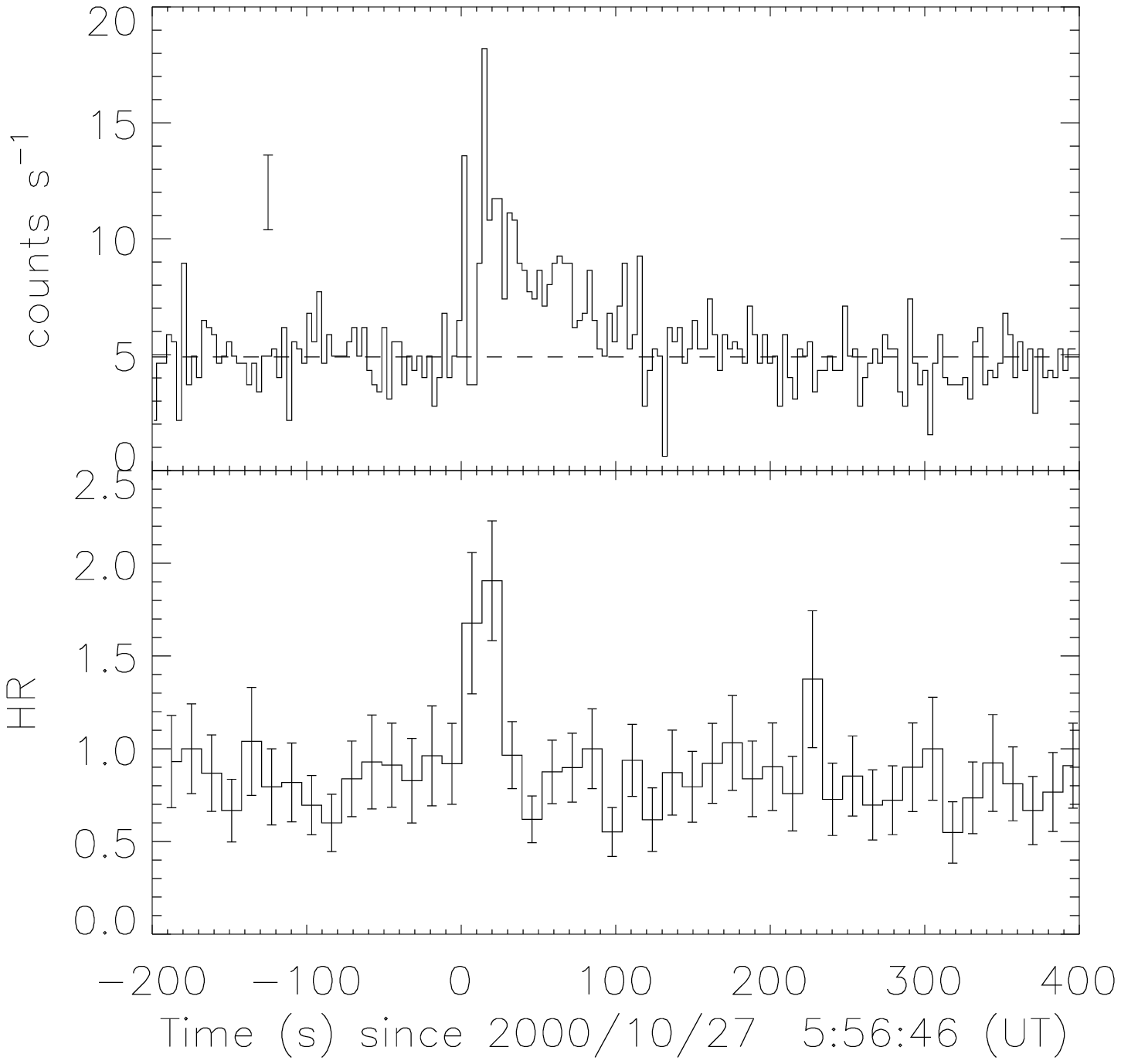,width=\linewidth}}
\caption{Light curve ({\it top panel}) and hardness ratio 
(4.5--8.0~keV:0.5--4.5~keV; {\it bottom panel}) from 600~s 
during which the 100~s burst was observed. The spectrum of the burst is harder
in the peak than in the tail, which is consistent with the cooling
that is expected in the tail of a thermonuclear X-ray burst. A typical 
uncertainty from the light curve is displayed in the upper left of the 
{\it top panel}.}
\label{fig:burst}
\end{inlinefigure}

\noindent
the persistent 
emission. For a 2~keV blackbody \citep[e.g.,][see also below]{kul02}, 
the peak rate of 18 count s$^{-1}$ is equivalent to 
$6 \times 10^{-10}$ erg cm$^{-2}$ s$^{-1}$, or $5 \times 10^{36}$~erg s$^{-1}$
at 8kpc (2--8 keV). The fluence of the burst is approximately 400 counts. 
%80 s 
%for a 2 keV blackbody and 1e23 of absorption, 1 c/s = 3.6e-11 erg cm^-2 s^-1
This translates to $1 \times 10^{-8}$~erg cm$^{-2}$ (2--8~keV, absorbed) 
assuming a 2~keV blackbody spectrum and a $1 \times 10^{23}$ cm$^{-2}$ 
absorption column, for an energy output of $10^{38}$~erg (8 kpc). 

We extracted a spectrum of the first 80 s of the burst in order to determine
the average temperature and solid angle of the emitting area. We used the 
persistent emission from the entire observation prior to 20~s before 
the burst as background \citep[e.g.,][]{kul02}, and modeled the burst as a 
blackbody
absorbed by a $10 \times 10^{22}$ cm$^{-2}$ column of gas and dust using
\program{XSPEC}. The 
spectrum contained only about 250 counts, so we grouped the spectrum so
that each bin contained $\ge 20$ counts. We can only place 90\% confidence
limits on the apparent temperature of 2--8~keV, and on the apparent radius of 
0.3--1.5~km (at 8~kpc). The fit was acceptable, with $\chi^2 = 6$ for 12 
degrees of freedom. The solid angles derived from 
spectra of X-ray bursts are usually within a factor of two of the expected
radius of a neutron star 
\citep[about 10~km; e.g.,][]{kap00,coc01,cor02a,kul02}.
The small solid angle from which the burst emission originates appears to be
the main factor in its relative weakness compared to previous bursts from
this source \citep[compare][]{coc99}.

In order to search for variability in the light curve aside from the burst, 
we removed 140 s of data starting 20 s prior to the burst, filled the 
gap with the mean count rate, and computed a Fourier Transform of the 
resulting light curve. We find broad-band low-frequency 
variability with a total rms power of 0.7\% rms above the expected Poisson 
noise between $5.5 \times 10^{-5}$ and 0.13~ Hz.
We find no evidence for 
periodic signals  over this frequency range with a significance
greater than 2-$\sigma$, and place a 90\% upper limit on any periodic signal 
in this range of 1.5\%, after incorporating the expected distribution 
of noise power \citep{vau94}. 
%% signal at 0.12857666 has power of 24.7, 97.5\% significant, not even 2-sigma

\subsection{\xmm}

\xmm\ observed \gcllb\ several times as part of a survey of the Galactic 
Plane and observations of the bursting pulsar GRO~J1744$-$28. We searched
for \gcllb\ in the three observations taken with the European Photon Imaging
Camera (EPIC) that were in the public archive as of April 2003 
(Table~\ref{tab:xmm}). 
Data from the two metal-oxide semiconductor 
(MOS) CCDs were available from all three observations, while data from the 
pn CCD camera were only available on 2001 September 04. The cameras record
X-rays within a calibrated energy band of 0.2--12~keV with an energy 
resolution of $E/\Delta E \approx 50$ at 6.5~keV.
The medium filters were used for all observations.

We reduced the data starting with the standard pipeline event lists for each 
EPIC camera using SAS version 5.4.1. We filtered the data to remove event 
grades higher than 12, 
events flagged as bad by the standard processing, and events below 0.3~keV 
(as these are likely to be background events). We then examined the light
curve from each observation in order to check for soft proton flares. The 
entire exposure on 2001 March 31 was contaminated with such flares, limiting
the usefulness of the data. Flares were also observed on 2001 September 04, 
so we removed all intervals for which the count rate was 3
standard deviations above the mean. We removed 6000~s of 
flaring from the pn data in this manner (resulting in the exposure listed
in Table~\ref{tab:xmm}), and 3500 s from the MOS data.

The field of view of the EPIC cameras are 30\arcmin\ in radius, so \gcllb\ 
was near the edge of the field in all three observations. \gcllb\ was
not detected by eye, and no source at its location was listed in the 
source catalogs that are produced in the standard pipeline processing. 
We were unable to use the technique applied to the \chandra\ observations
to derive an upper limit on the flux from \gcllb\ during the \xmm\ 
observations for two reasons. First, although the full-width of the 
half-maximum of the \xmm\ PSF is only 6\arcsec\ on-axis, off-axis it is 
significantly degraded. At an offset of 
10\arcmin, 90\% of the photons at 5~keV are enclosed by the PSF within 
radii of 70\arcsec\ for the pn CCDs and 85\arcsec\ for the MOS CCDs. Second,
the diffuse X-ray emission detected with \chandra\ from the ``darkest'' 
regions of the inner 10\arcmin\ of the Galaxy varies by 10\% around a mean 
value of $2\times 10^{-13}$~\ergcms arcmin$^{-2}$ 
(Muno et al., in preparation). 
We therefore used the fluxes from the faintest sources detected as part of the 
pipeline processing between 8\arcmin\ and 12\arcmin\ from the aim-point of 
each observation as estimates of the upper limit on the count rate from
\gcllb. These count rates were 
0.009 count s$^{-1}$ (MOS, 2--12~keV) for 2001 March 31,
0.003 count s$^{-1}$ (MOS, 2--12~keV) for 2001 April 01, and
0.002 count s$^{-1}$ 

\begin{inlinefigure}
\centerline{\epsfig{file=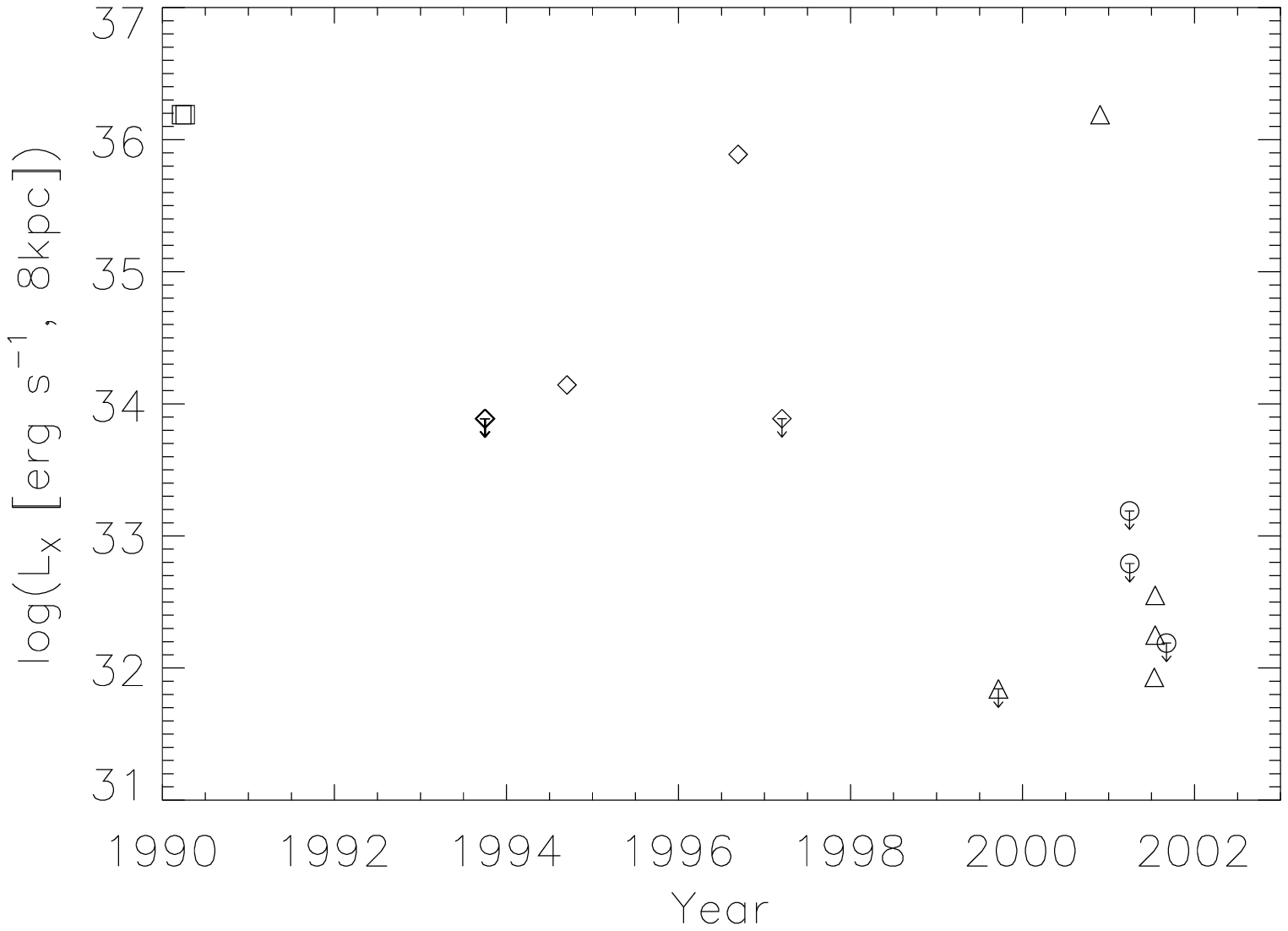,width=\linewidth}}
\caption{The luminosity history of \gcllb\ assembled from \granat\ (squares;
3--30~keV), 
\asca\ (diamonds; 2--10~keV), 
\xmm\ (circles; 2--12~keV), 
and \chandra\ (triangles; 2--8~keV) observations. 
The luminosity varies over four orders
of magnitude on time scales of months, as is commonly observed from 
LMXB transients. The variability at low luminosities 
($< 10^{33}$~\ergsec) is significant (see Table~3 for uncertainties), and 
has been observed from other sources on only a few occasions.}
\label{fig:fluxhistory}
\end{inlinefigure}

\noindent
(pn, 2--12~keV) for 2001 September 04.
We converted these to energy fluxes using \program{PIMMS} by 
assuming a $\Gamma=1.9$ power law and $N_{\rm H} = 10\times10^{22}$~cm$^{-2}$, 
and listed them in Table~\ref{tab:fluxhistory}. The resulting limits are
similar to both the \chandra\ detections in quiescence and the 
variations in the background flux on arcminute scales.

\section{Discussion}

We plot the long-term history of the persistent luminosity from \gcllb\
in Figure~\ref{fig:fluxhistory} (deabsorbed, and assuming a distance of 
8~kpc). It is clear that this source varies
in luminosity by at least four orders of magnitude on time scales of 
months. \gcllb\ was fainter than $10^{33}$~\ergsec in Fall 1999, and Fall 
and Spring 2001. The source has been observed to reach luminosities
above $10^{36}$~\ergsec\ three times in the last 13 years: in Spring 1990,
Fall 1996, and Fall 2001. This suggests that the bright states are transient 
outbursts driven by instabilities in the accretion disk 
\citep[e.g.,][]{las01}. 
It is unlikely that there were any outbursts brighter than 
$6 \times 10^{36}$~\ergsec\ lasting more than a month 
between 1996 and the present, because they would have been detectable with the 
\bepposax\ WFC and the \rxte\ ASM. Thus, \gcllb\ appears to be a genuine 
low-luminosity transient.
Under the disk instability model, the faintness of the outbursts can be 
explained if the mass transfer rate
from the companion is low, either because it is due to a weak wind, or
because the companion is a very low-mass star that fills its Roche-lobe
\citep{king00}.

\gcllb\ was also detected several times at luminosities below 
$10^{35}$~\ergsec\ (Figure~\ref{fig:fluxhistory}). 
It was observed with 
\asca\ at a luminosity of $10^{34}$~\ergsec\ Fall 1994. Several other
sources have been observed with varying flux at this level, including
the neutron star 
LMXBs SAX~J1747.0$-$2853 \citep{wij02b}, SAX~J1808.4$-$3568 \citep{wij01}, 
1RXS~J1718.2$-$402934 \citep{kap00}, and SAX~J1828.5$-$1037~\citep{cor02a}.   
Such low-luminosity activity is not addressed by the simplest disk 
instability models, which generally produce $10^{37} - 10^{38}$~\ergsec\ 
outbursts because a large fraction of the accretion disk is expected to be 
disrupted (e.g., Dubus, Hameury, \& Lasota 2001). Furthermore,
\chandra\ observations demonstrate that \gcllb\ varies in X-ray intensity 
when its luminosity is only about $10^{32}$~\ergsec (Table~\ref{tab:counts}
and Figure~\ref{fig:fluxhistory}). The fact that \gcllb\ is significantly
brighter after the Fall 2000 outburst could be attributed to the flux 
from the hot ($kT \sim 0.3$ keV) surface of the neutron star, which may 
have been heated during the outburst \citep{bbr98}. However, the high 
absorption column toward the source should make most of this thermal 
emission unobservable (see Figure~\ref{fig:spec}). On the other
hand, short time-scale variations have been observed in the flux from the 
LMXBs Cen~X-4 \citep{cam97,rut01} and Aql~X-1 \citep{rut02}
that are inconsistent with a cooling neutron star. This emission
has been explained as residual accretion either onto the neutron star 
surface or onto the magnetospheric boundary \citep{cam98}.
The emission from \gcllb\ with luminosities between
$10^{32}$ and $10^{34}$~\ergsec\ probably also represents continued  
accretion. These low luminosities are an important, and relatively unexplored, 
regime of accretion that challenge current disk instability models 
for LMXBs.

X-ray bursts were observed from \gcllb\ with the \bepposax\ WFC during 
the bright 
state in Fall 1996 \citep{coc99,sak02}, and with \chandra\ during the 
outburst in Fall 2000 (this work). The WFC observed the Galactic Center
for approximately 500~ks each spring and fall from mid-1996 through the end 
of 2001,
and detected no further bursts. This suggests that the bursts only 
occur frequently when the persistent X-ray emission reaches $\sim 10^{36}$ 
\ergsec, so that
accretion provides $10^{-10}$ \msunyr\ of fuel to the surface of the neutron 
star. It is likely that some other low-luminosity bursters are similar,
and that sensitive observations at or just before the times that bursts 
were observed would have revealed persistent emission just below the 
sensitivity limit of the \bepposax\ WFC and \rxte\ ASM 
\citep[$\sim 10^{36}$~\ergsec; see also][]{coc01}.
If this is the case, then the fraction of low-luminosity bursters that
have been detected so far depends upon how often the transient outbursts
occur, which is currently uncertain.
%[No other low-luminosity bursters are in the \asca\ survey.]

%SAX J1752.3$-$3138 \citep{coc01} X-ray bursts often seen after transient 
%activity has ended. XTE J1709$-$276, SAX J1712.6$-$3739, SAX J1750.8-2900,
%SAX J1810.6-2609, SAX J1808.4-3658
%in 't Zand, J. J. M. 2000, Proc. 4th INTEGRAL Worskshop (astro-ph/0104299)
%(didn't find that stated clearly)

The weakness of the X-ray burst observed with \chandra\ is a bit 
puzzling. Its peak flux of $6 \times 10^{-10}$~\ergcms\ (2--8~keV) 
is a factor of 40 below those of Eddington-limited bursts observed 
previously from this source with the \bepposax\ WFC ($10^{-8}$~\ergcms).
Faint bursts are not uncommon from LMXBs, but they are usually preceded 
by bright bursts tens of minutes before \citep[e.g.][]{lan87,got86}, and
are thought to occur in fuel that was not burnt in the 
previous burst \citep{lew87, fuj87}. However, no burst was observed 
from \gcllb\ in the seven hours of \chandra\ observations prior to the 
weak burst, and the fluence of the burst ($10^{38}$~erg) is an order of 
magnitude smaller than the amount
of nuclear energy that could have been stored in accreted hydrogen
during the \chandra\ observation prior to the burst
($0.04 \cdot 2.5\times10^{4}~{\rm s} \cdot 10^{36}~{\rm erg~s}^{-1} = 
10^{39}~{\rm erg}$; see Figure~\ref{fig:lc}). 
%
% Assume H eddington limit for 10 km, 1.4 solar mass NS of 1.3e38 erg/s
% 2e36 erg/s => 2e-10 Msun yr^-1 = 1e16g/s = 0.015 L_edd, log(L/Ledd) = -1.8
% just on border between pure He bursts and unstable H-triggered bursts.
% At these accretion rates, the critical surface density at which a burst
% ignites is $10^9 - 10^{10}$ g cm$^{-1}$ 
%\citep[][we assume a 10 km, 1.4 M$_odot$ star]{nh03}. 
% Assuming the matter is accreted
% and burned on the entire surface of the neutron star, the total amount
% of fuel available for a burst is $10^{22} - 10^{33}$ g, which would 
% contain a nuclear energy of $10^{41} - 10^{42}$ erg (assuming hydrogen 
% fuel with 8 MeV per nucleon, or $8 \times 10^{18}$ erg g$^{-1}$).
%
%Weak bursts are seen from edge-on systems in 
%which the accretion disk obscures the neutron star, although the lack of 
%periodic X-ray modulations over 10 hours in \gcllb\ argues against this 
%interpretation. 
On the other hand, this weak burst is reminiscent of a $10^{37}$~erg burst 
observed by \citet{gk97} with \asca\ from the globular cluster M28. 
The source that produced that burst had a persistent luminosity 
of at most $10^{33}$~\ergsec. However, if material was accreted onto the 
entire surface of the neutron star at such a low accretion rate 
($< 10^{-13}$ \msunyr), the recurrence time for bursts should be very long, 
and their fluence high. For instance, to collect the $10^8$ g cm$^{-2}$ column 
necessary to trigger a burst over the entire star (an area of $10^{13}$ 
cm$^{-2}$) would take several years, and the nuclear energy stored on the 
surface would be at least $10^{39}$ erg \citep[e.g.,][]{fhm81}. 
Therefore, \citet{gk97} suggested that the accreted material in the M28 source
is channeled onto a small area of the surface of the neutron star, such
as its magnetic poles. This 
explanation may also apply for \gcllb, for which we infer that the burst
emission originates from a solid angle that is only a few percent of the 
surface area of a 10 km neutron star. 
Moreover, the fact that four low-luminosity 
%out of about 15 
transients are accreting millisecond X-ray pulsars suggests that faint 
transients are 
more likely than the general LMXB population to have magnetic fields strong 
enough to channel the accretion flow onto the poles of the neutron star. 
Therefore, we suggest that future observations of \gcllb\ in outburst 
with sensitive timing instruments could reveal millisecond pulsations.

%Unfortunately, confirmation that this is the case in \gcllb\ will have to 
%await the next generation of X-ray instruments. The signals from accreting 
%millisecond pulsars typically have amplitudes $\lessim 10$\%, and their
%frequencies vary on time scales of hundreds of seconds due to the Doppler 
%motion of the primary. Current instruments are therefore not sensitive 
%enough to detect a pulsar signal in a faint, $10^{-10}$~\ergcms\ source 
%such as \gcllb\ in outburst.

\acknowledgements{We thank A. Levine, D. Galloway, R. Remillard, and C.
Markwardt for useful discussions, and for providing information on the 
flux history and burst properties of \gcllb\ and similar LMXBs. We thank
the referee for the careful reading and useful comments.
This work has been supported by NASA grants NAS 8-39073 and NAS 8-00128.}

\clearpage

\begin{deluxetable}{lccc}
%\tabletypesize{\scriptsize}
\tablecolumns{4}
\tablewidth{0pc}
\tablecaption{Flux History of \gcllb\label{tab:fluxhistory}} 
\tablehead{
\colhead{Date} & \colhead{Mission} & \colhead{$F_{\rm X}$\tablenotemark{a}} &
\colhead{References} 
} \startdata
1990 Mar 24 & {\it Granat} & 2000 & (1) \\
1990 Apr 08 & {\it Granat} & 2000 & (1) \\
1993 Sep 30 & {\it ASCA} & $<$12 & (2) \\
1993 Oct 01 & {\it ASCA} & $<$40 & (2) \\
1993 Oct 03 & {\it ASCA} & $<$20 & (2) \\
1993 Oct 04 & {\it ASCA} & $<$19 & (2) \\
1994 Sep 15 & {\it ASCA} & 18 & (2) \\
1996 Sep 11 & {\it ASCA} & 1000 & (2) \\
1997 Mar 16 & {\it ASCA} & $<$8 & (2) \\
1999 Sep 21 & {\it Chandra} & $<$0.09 & (3) \\
2000 Oct 26 & {\it Chandra} & 2000 & (3) \\
2001 Mar 31 & {\it XMM} & $<$2 & (3) \\
2001 Apr 01 & {\it XMM} & $<$0.8 & (3) \\
2001 Jul 14 & {\it Chandra} & 0.1 & (3) \\
2001 Jul 18 & {\it Chandra} & 0.3 & (3) \\
2001 Jul 18 & {\it Chandra} & 0.5 & (3) \\
2001 Sep 04 & {\it XMM} & $<$0.2 & (3) \\
\enddata
\tablenotetext{a}{X-ray flux in units of $10^{-13}$~ erg cm$^{-2}$ s$^{-1}$.
The energy ranges are: 3--30~keV for {\it Granat}, 2--10~keV for {\it ASCA},
2--8~keV for {\it Chandra}, and 2--12~keV for {\it XMM}.}
\tablecomments{The {\it BeppoSAX} WFC observed the Galactic center every Spring 
and Fall from mid-1996 through the end of 2001. The WFC detected three 
thermonuclear bursts during Fall 1996, but never detected persistent emission 
from accretion with $F_{\rm X} > 2 \times 10^{-10}$~erg cm$^{-1}$ s$^{-1}$
\citep{coc99}.}
\tablerefs{(1) \citet{pav94}; (2) \citet{sak02}; (3) This work.}
\end{deluxetable}

% RXTE PCA scans sensitive to 1e-11 erg cm^-2 s^-2; \gcllb\ is 
% awfully close to XTE 1748-288, SAX J1747-2853, and Sgr A*
% 1 Crab is 2.8e-8 erg cm^-2 s^-1, 2-12 keV.

\begin{deluxetable}{lcccccc}
%\tabletypesize{\scriptsize}
\tablecolumns{7}
\tablewidth{0pc}
\tablecaption{\chandra\ Observations of \gcllb\label{tab:obs}}
\tablehead{
\colhead{} & \colhead{} & \colhead{} & 
\multicolumn{2}{c}{Aim Point} & \colhead{}  & \colhead{} \\
\colhead{Start Time} & \colhead{ObsID} & \colhead{Exposure} & 
\colhead{RA} & \colhead{DEC} & \colhead{Roll} & \colhead{Offset\tablenotemark{a}} \\
\colhead{(UT)} & \colhead{} & \colhead{(s)} 
& \multicolumn{2}{c}{(degrees J2000)} & \colhead{(degrees)} & \colhead{(arcmin)}
} \startdata
1999 Sep 21 02:43:00 & 0242  & 40,872 & 266.41382 & -29.0130 & 268 & 10 \\
2000 Oct 26 18:15:11 & 1561a\tablenotemark{b} & 35,705 & 266.41344 & -29.0128 & 265 & 10 \\
2001 Jul 14 01:51:10 & 1561b & 13,504 & 266.41344 & -29.0128 & 265 & 10 \\
2001 Jul 18 14:25:48 & 2284 & 10,625 & 266.40415 & -28.9409 & 284 & 8 \\
2001 Jul 18 17:37:38 & 2287 & 10,625 & 266.21412 & -28.8391 & 284 & 5 \\
\enddata
\tablenotetext{a}{The offset listed is from the nominal position of 
\gcllb\ from \citet{pav94}. }
\tablenotetext{b}{This observation was taken during two epochs because the 
ACIS-I camera was powered off at the beginning of the first epoch. 
It is listed in the archive under
a single ObsID, 1561.}
\end{deluxetable}

\begin{deluxetable}{lcccc}
%\tabletypesize{\scriptsize}
\tablecolumns{5}
\tablewidth{0pc}
\tablecaption{\chandra\ Photometry of \gcllb\ (2--8~keV)\label{tab:counts}}
\tablehead{
\colhead{} & \multicolumn{3}{c}{Counts} & \colhead{}\\
\colhead{Date} & \colhead{Tot} & \colhead{Bkgd} & 
\colhead{Net} & \colhead{Flux}
} \startdata
1999 Sep 21 02:43 & 30 & 27.6 & $<11.9$ & $<0.09$\\
2000 Oct 26 18:15 & 170214 & 87 & 170127$\pm$412 & $2000\pm 120$ \\
2001 Jul 14 01:51 & 14 & 8.6 & 5.4$^{+6.6}_{-5.0}$ & $0.11^{+0.15}_{-0.10}$ \\
2001 Jul 18 14:26 & 11 & 2.0 & 9.0$^{+6.4}_{-4.6}$ & $0.23^{+0.19}_{-0.13}$ \\
2001 Jul 18 17:38 & 18 & 0.8 & 17.2$^{+7.9}_{-6.1}$ & $0.46^{+0.23}_{-0.18}$ \\
\enddata
\tablecomments{Confidence limits are 90\%, based on the observed number of 
counts (see text). 
Fluxes are absorbed in units of $10^{-13}$~\ergcms\ (2--8~keV).}
\end{deluxetable}
% Gamma = 1.2 power law, NH = 8.5e22:  1e-5 c/s = 3.1e-16 erg cm^-2 s^-1
%					unabsorbed = 5.2e-16
% kT=0.3 bb 					= 1.1e-16 erg cm^-2 s^-1
%					unabsorbed = 9.0e-16
% assuming that the ocunt rate is 17.2 count/10 ks, what is chance of 
% getting <3.1 counts randomly?

\begin{deluxetable}{lcccccc}
%\tabletypesize{\scriptsize}
\tablecolumns{7}
\tablewidth{0pc}
\tablecaption{\xmm\ Observations of \gcllb\label{tab:xmm}}
\tablehead{
\colhead{} & \colhead{} & \colhead{} & 
\multicolumn{2}{c}{Aim Point} & \colhead{}  & \colhead{} \\
\colhead{Start Time} & \colhead{Sequence} & \colhead{Exposure} & 
\colhead{RA} & \colhead{DEC} & \colhead{Roll} & \colhead{Offset} \\
\colhead{(UT)} & \colhead{} & \colhead{(s)} 
& \multicolumn{2}{c}{(degrees J2000)} & \colhead{(degrees)} & \colhead{(arcmin)}
} \startdata
2001 Mar 31 11:24:26 & 0112971601 & 3830 & 266.40146 & $-$28.9820 & 95 & 10 \\
2001 Apr 01 13:02:43 & 0112971901 & 9191 & 266.12667 & $-$28.7313 & 95 & 10 \\
2001 Sep 04 01:19:34 & 0112972101 & 19,040 & 266.43463 & $-$29.0313 & 271 & 10  \\
\enddata
\tablecomments{The offset listed is from the nominal position of \gcllb\ from
\citet{pav94}. A fourth observation, 0112972001, is listed in the archive, 
but contains no usable data.}
\end{deluxetable}
%
%
%2101: 22540 for MOS, 26050 without flares

\end{document}